\documentclass[pdflatex]{article}
\usepackage{dcase2019,amsmath,times}
\usepackage{graphicx}
\usepackage{bm}
\usepackage{url}
\usepackage{amssymb}
\usepackage{cite}
\makeatletter
\def\Hline{
  \noalign{\ifnum0=`}\fi\hrule \@height 2.\arrayrulewidth \futurelet
  \reserved@a\@xhline}
\makeatother

\title{DOA Estimation by DNN-based Denoising and Dereverberation from Sound Intensity Vector}

 \name{Masahiro Yasuda${}^{1}$, Yuma Koizumi${}^{1}$, Luca Mazzon${}^{2}$, Shoichiro Saito${}^{1}$and Hisashi Uematsu${}^{1}$}
\address{
${}^{1}$NTT Media Intelligence Laboratories, Tokyo, Japan\\
${}^{2}$University of Padova, Padua, Italy
}
\begin{document}

\ninept
\maketitle

\begin{sloppy}

\begin{abstract}
We propose a direction of arrival (DOA) estimation method that combines sound-intensity vector (IV)-based DOA estimation and DNN-based denoising and dereverberation. Since the accuracy of IV-based DOA estimation degrades due to environmental noise and reverberation, two DNNs are used to remove such effects from the observed IVs. DOA is then estimated from the refined IVs based on the physics of wave propagation. Experiments on an open dataset showed that the average DOA error of the proposed method was 0.528 degrees, and it outperformed a conventional IV-based and DNN-based DOA estimation method.
\end{abstract}

\section{Introduction}
\label{sec:intro}
Time series direction-of-arrival (DOA) estimation, which is  the  task of identifying the relative position of  the sound sources with respect to the microphone at every time frame, is an important technology for understanding the surrounding environment from sound recordings. For example, DOA estimation is useful for autonomous driving that autonomously acquiring the surrounding environment~\cite{SmartCar1}. DOA estimation is also used as a component of surveillance systems via the microphone array carried in a drone~\cite{drone}.

Recent DOA estimation methods can be broadly classified into two categories: parametric based \cite{GCC-PHAT,MUSIC,DOA_FOAIV} and machine-learning based \cite{SELD-DOA,SELD-DOA2}. Various parametric-based methods have been proposed, such as a method based on time difference of arrival (TDOA), e.g., generalized cross correlation with phase transform (GCC-PHAT) ~\cite{GCC-PHAT} and a subspace method, e.g., multiple-signal-classification (MUSIC) \cite{MUSIC}. Using a deep neural network (DNN) is a recent advancement in machine-learning-based methods. Several methods have been proposed that use a DNN as a regression function for directly estimating DOA from observed signals~\cite{SELD-DOA,SELD-DOA2}.

Both parametric-based and DNN-based methods have advantages and disadvantages. Parametric-based methods can accurately estimate DOAs when the maximum number of sources is known. However, since these methods use many time-frames for DOA estimation, there is a trade-off relationship between the accuracy of time-series analysis and angle estimation. DOA estimation using sound intensity vectors (IVs)~\cite{DOA_FOAIV} allows time-series analysis with good time-angular resolution. However, its accuracy is affected by the signal-to-noise ratio (SNR) corresponding to environmental noise and reverberation. On the other hand, DNN-based DOA estimation methods are robust against SNR \cite{Kapka_DOA,Cao_DOA}. However, conventional end-to-end approaches cannot combine physical knowledge of wave propagation because DNN-based DOA estimation methods' processing mechanisms are black boxes.

\begin{figure}[t]
  \centering
  \includegraphics[width=80mm,clip]{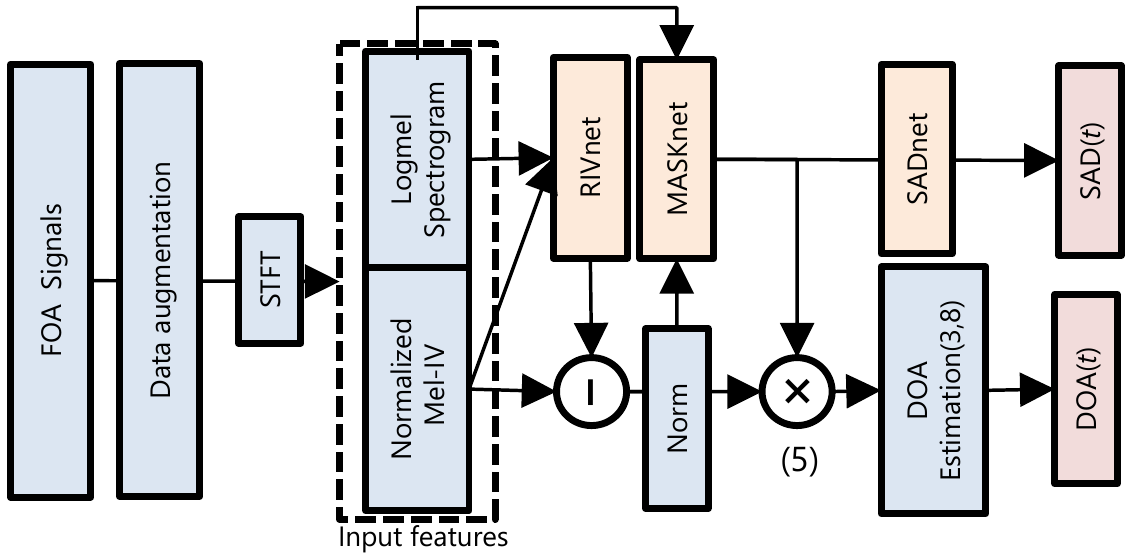}
  \caption{System overview.}
  \label{fig:overview}
\end{figure}

We propose a time series DOA estimation method that combines the advantages of parametric-based and DNN-based methods: an IV-based method with the first order ambisonics (FOA) format signal is used as the parametric-based methodbased method , and two DNNs assist it by removing environmental effects such as DNN-based denoising, as shown in Fig. \ref{fig:overview}. One of the DNNs, called MASKnet, works to reduce noise by multiplying a time-frequency (T-F) mask, and the other DNN, called RIVnet, works to subtract other effects that cannot be removed by mask-based denoising such as reverberation.

\section{CONVENTIONAL METHODS}
\label{sec:conv}
\subsection{DOA estimation using intensity vector}
\label{sec:iv_based}
Ahonen {\it et al.} proposed a DOA estimation method using IVs calculated from a set of FOA B-format recordings \cite{DOA_FOAIV}. The FOA B-format consists of four channels of signals, and its short-time Fourier transform (STFT) outputs ${\rm W}_{f,t},{\rm X}_{f,t},{\rm Y}_{f,t}$, and ${\rm Z}_{f,t}$ corresponding to the 0-th and 1st order of spherical harmonics. Here, $f \in \{1,...,F\}$ and $t \in \{1,...,T\}$ are indexes of frequency and time-frame in the time-frequency (T-F) domain, respectively. The 0-th harmonic ${\rm W}_{f,t}$ corresponds to a non-directional sound source, and the 1st harmonics ${\rm X}_{f,t}$, ${\rm Y}_{f,t}$, and ${\rm Z}_{f,t}$ correspond to the dipoles along each axis, respectively. The spatial responses (steering vectors) of ${\rm W}_{f,t},{\rm X}_{f,t},{\rm Y}_{f,t}$, and ${\rm Z}_{f,t}$ are defined as $H^{(W)} (\phi,\theta,f) = 3^{-1/2}$, $H^{(X)}(\phi,\theta,f) = \cos\phi\ast\cos\theta$, $H^{(Y)}(\phi,\theta,f) = \sin\phi\ast\cos\theta$, and $H^{(Z)}(\phi,\theta,f) = \sin\theta$ respectively. Here, $\phi$ and $\theta$ are the azimuth and elevation angle,respectively. 

Originally, an IV is defined in the T-F domain as ${\bf I}_{f,t} = \frac{1}{2}\mathfrak{R} (p_{f,t}^*\cdot{\bf v}_{f,t} )$, where ${\bf v} = [v_x,v_y,v_z]^{\top}$ is the sound-particle velocity, $p_{f,t}$ is the sound pressure in the T-F space, $\mathfrak{R} (\cdot)$ denotes the real-part of complex numbers, and ${}^*$ is the conjugate of complex numbers. Since it is impossible to measure sound pressure and sound velocity at continuous points, calculation of ${\bf I}_{f,t}$ is difficult. As an approximation, the IV of each T-F bin can be calculated from the 4-channel spectrograms of the FOA B-format as
\begin{equation}
    {\bf I}_{f,t}\propto \mathfrak{R}\left({\rm W^*}_{f,t}
    \bm{h}_{f,t}
    \right)
    = 
    \left[ 
    I_{X,f,t},
    I_{Y,f,t},
    I_{Z,f,t}
    \right]^{\top},
    \label{eq:int}
\end{equation}
where $\bm{h}_{f,t} = \left[ {\rm X}_{f,t}, {\rm Y}_{f,t}, {\rm Z}_{f,t} \right]^{\top}$. To select an effective T-F domain, Ahonen {\it et al.} \cite{DOA_FOAIV} applied T-F Mask $M_{f,t}$ to the IV spectrogram. The mask is defined as
\begin{equation}
        M_{f,t} = \lambda \left(|{\rm W}_{f,t}|^2 + \frac{|{\rm X}_{f,t}|^2+|{\rm Y}_{f,t}|^2+|{\rm Z}_{f,t}|^2}{3}\right),
\end{equation}
where $\lambda = (2\rho_0c^2)^{-1}$. This mask has a high -value at the large-power T-F bin, and a low value at the small-power T-F bin. By assuming that the target source has higher power than environmental noise, the T-F mask selects an effective T-F region for DOA estimation of the target source. It then sums the IVs for all frequencies at each time-frame and obtain time-series IVs. Finally, DOA of the target source is estimated in each time frame $t$ as
\begin{equation}
\phi_t = \arctan\left(\frac{I_{Y,t}}{I_{X,t}}\right), \;\;\; \theta_t = \arctan\left(\frac{I_{Z,t}}{\sqrt{I_{X,t}^2 + I_{Y,t}^2}}\right).
\label{eq:extract_doa}
\end{equation}

\subsection{DNN-based methods}
A recent advancement in DOA estimation is the use of a DNN as a regression function for directly estimating the azimuth and elevation labels from observations \cite{SELD-DOA,SELD-DOA2,Kapka_DOA,Cao_DOA}. Several DNN-based methods outperform conventional parametric DOA estimation methods without the need of any physical knowledge, that is, perfectly data-driven approach. In fact, many participants of an international technical competition of DOA estimation\footnote{Task 3 of the IEEE AASP Challenge on Detection and Classification of Acoustic Scenes and Events (DCASE)} used perfectly data-driven approaches \cite{Kapka_DOA,Cao_DOA} and achieved good accuracy. In these approaches, the DNN structure is a convolutional recurrent neural network (CRNN) that is combination of a multi-layer convolutional neural network (CNN) and bidirectional-gated recurrent units (Bi-GRUs), which enable extraction of higher-order features and modeling of temporal structure, and the DNN was trained to minimize the metric, such as the mean-absolute error (MAE), between the true and estimated DOA.

\section{PROPOSED METHOD}
\label{sec:proposed}
\subsection{Basic concept}
Our DOA estimation method  uses IVs refined using both a T-F mask and reverberation components estimated using DNNs. Generally, a time-domain input signal can be expressed as the sum of the components of direct sound, reverberation, and noise. According to this modeling, its T-F representation can also be written as the sum of these components. Thus, the IV calculated using (\ref{eq:int}) can be expressed~as
\begin{equation}
    {\bf I}_{f,t} = {\bf I}^s_{f,t} + {\bf I}^r_{f,t} + {\bf I}^n_{f,t},
    \label{eq:int3}
\end{equation}
where ${\bf I}^s_{f,t}$, ${\bf I}^r_{f,t}$, and ${\bf I}^n_{f,t}$ are the IVs of direct sound, reverberation, and noise, respectively. That is, time-series IVs ${\bf I}_t$ are affected by not only the direct sound but also reverberation and noise. This is one of the reasons conventional IV-based methods are not robust against reverberation and noise. 

To overcome this problem, we subtract the estimated reverberation component $\hat{{\bf I}}^r_{f,t}$ from ${\bf I}_{f,t}$ for dereverberation and multiply a T-F mask $ M_{f,t}$ by the obtained IVs for denoising. This is because the noise has little overlap in the T-F domain with the direct sound and can be removed with the T-F mask, but the reverberation is not.
This process can be written as
\begin{equation}
{\bf I}^s_{t} = \sum_{f}M_{f,t} \left( {\bf I}_{f,t} - \hat{{\bf I}}^r_{f,t} \right).
    \label{eq:int_s}
\end{equation}
We estimate $M_{f,t}$ and $\hat{{\bf I}}^r_{t,f}$ by using two DNNs, as shown in Fig. \ref{fig:overview}.


\begin{figure}[t]
  \centering
  \includegraphics[width=80mm,clip]{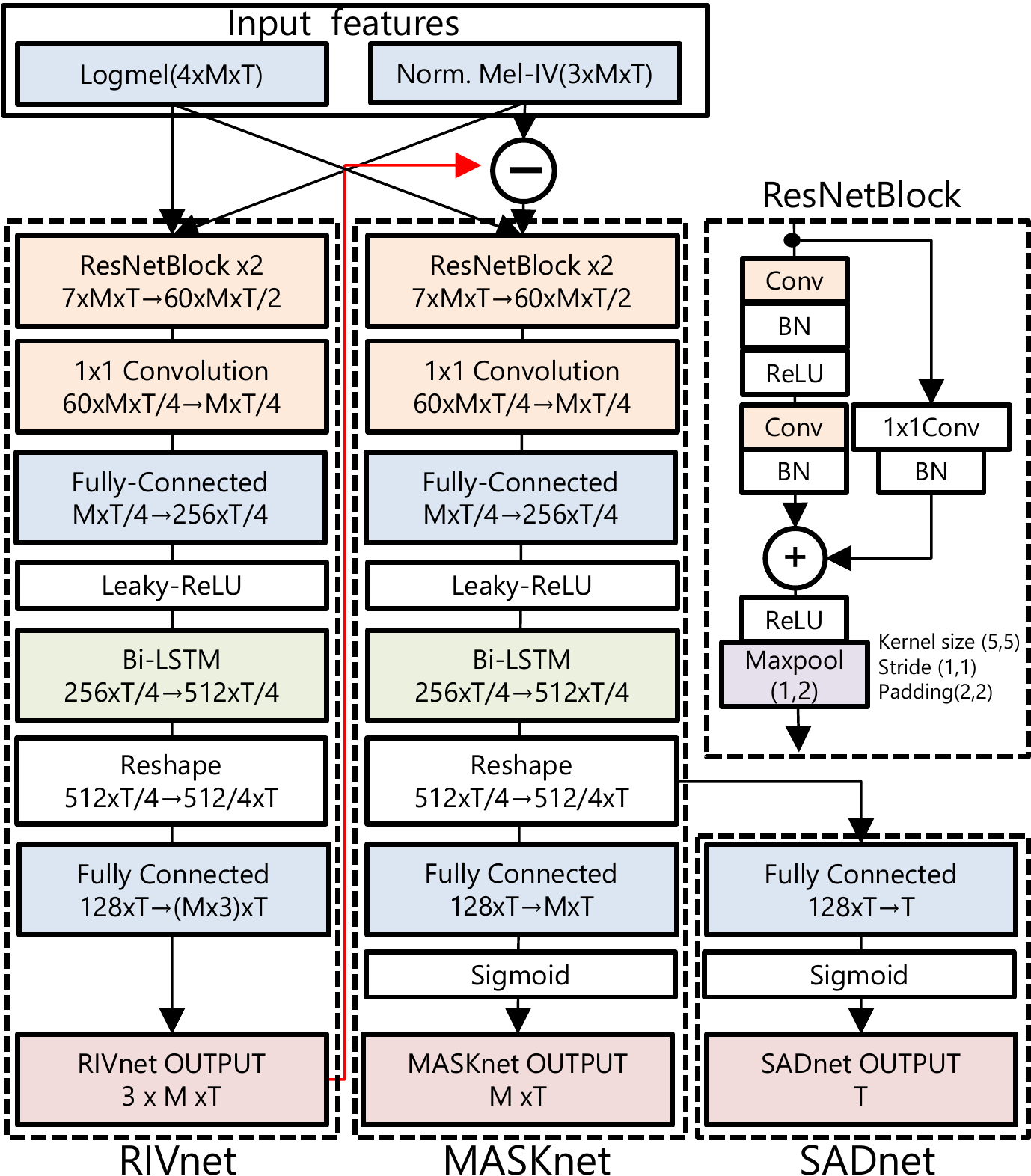}
  \caption{DNN architecture of the proposed method. In the figure of 
  ResNetBlock, ``Conv'', ``BN'', and ``Maxpool'' denotes convolutional layer, batch normalization, and max pooling, respectively.}
  \label{fig:DNNmodel}
\end{figure}
\subsection{Network architecture and loss function}
\subsubsection{Input features}
Figure \ref{fig:DNNmodel} shows an overview of the DNN architecture of the proposed method. First, IV ${\bf I}_{f,t}$ and logmel-spectrograms are extracted from the input signals. Note that IVs are also compressed by the Mel-filterbank to guarantee that their dimensions are the same as that of the logmel-spectrograms, as in a previous study~\cite{Cao_DOA}. This allows us to concatenate the IVs and logmel-spectrograms as an input feature for the first CNN layer of RIVnet. The IVs are also normalized as ${\bf I}^{\rm norm}_{f,t} = {\bf I}_{f,t}/|{\bf I}_{f,t}|$ because only the direction of a IV is necessary for DOA estimation by (\ref{eq:extract_doa}). RIVnet then estimates the reverberant components of IV $\hat{{\bf I}}^r_{f,t}$, and refined IV ${\bf I}'_{t,f}={\bf I}_{t,f}-\hat{{\bf I}}^r_{f,t}$ is estimated. Note that ${\bf I}'_{t,f}$ is also normalized. Then, ${\bf I}'_{t,f}$ and logmel-spectrograms are input to MASKnet to estimate a T-F mask. For sound-activity detection, we use a short branch of MASKnet with the sigmoid activation called SADnet. RIVnet and MASKnet multi-layer CNN blocks for high-level feature extraction and an RNN layer for modeling temporal structures. The final estimates of the azimuth and elevation are calculated using (\ref{eq:extract_doa}) from the IVs, which are refined by RIVnet output ${\bf \hat{I}}^r_{t,f}$ and MASKnet output $M_{t,f}$ as (\ref{eq:int_s}). The total number of trainable parameters is 2.79M.

\begin{figure*}[t]
  \centering
  \includegraphics[angle=90,width=175mm]{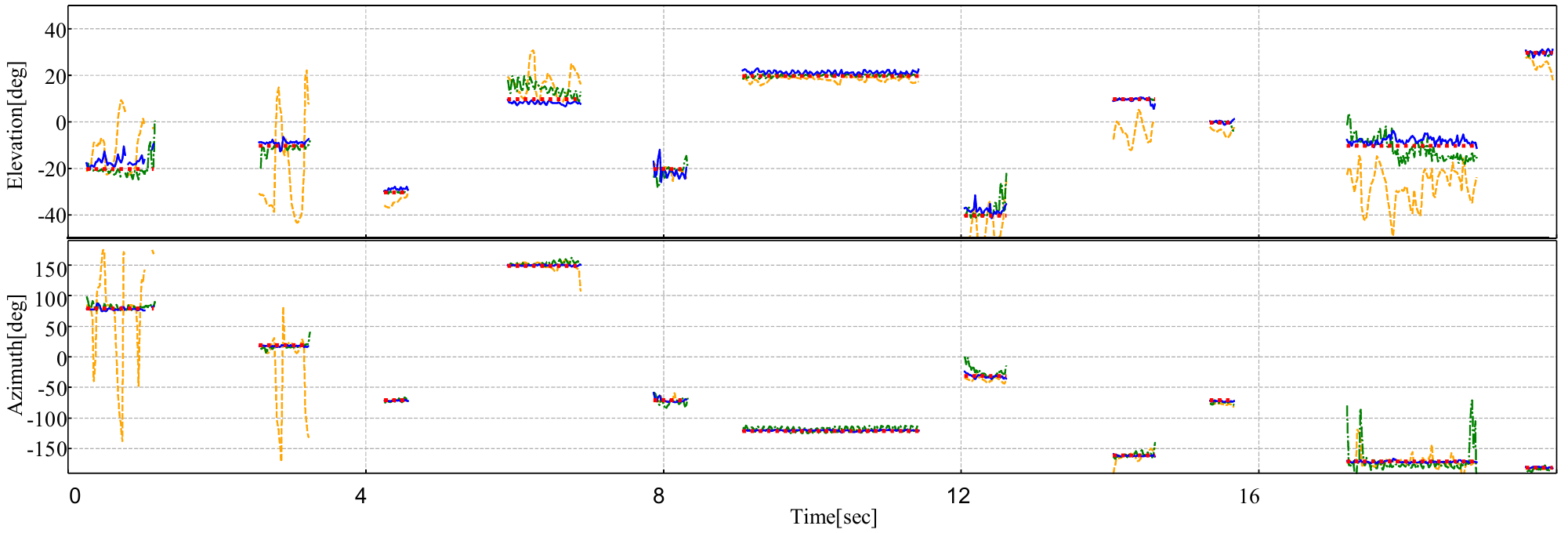}
  \caption{Example results of DOA estimation. Red dotted line represents ground-truth DOAs, orange dashed line represents estimated DOAs using IV-based method, green dash-dotted line represents estimated DOAs using conventional DNN-based method, and blue solid line represents estimated DOAs using MASK+RIV+SAD  with augmentation model.}
  \label{fig:DOAresult}
\end{figure*}

\begin{table*}[t]
\centering
\caption{Experimental results. FR denotes frame-recall.} 
\begin{tabular}{l|cc|cc|cc|cc|cc} \hline
&\multicolumn{2}{c}{Average}&\multicolumn{2}{|c|}{FOLD1}&\multicolumn{2}{c|}{FOLD2 }&\multicolumn{2}{c|}{FOLD3}&\multicolumn{2}{c}{FOLD4} \\ 
& DE & FR & DE & FR & DE & FR & DE & FR & DE & FR\\ \hline
IV-based \cite{DOA_FOAIV} & $10.5^\circ$ & - & $10.3^\circ$ & - &$9.72^\circ$ & - & $10.9^\circ$ &- & $11.2^\circ$&- \\ 
DNN-conv 
& $4.506^\circ$ & 0.949
& $5.343^\circ$ & 0.936 
& $2.848^\circ$ & 0.927 
& $5.045^\circ$ & {\bf 0.979}
& $4.787^\circ$ & 0.952 \\ \hline 
A: MASK+SAD &$1.29^\circ$ & 0.958 & $1.028^\circ$ & 0.959 &$1.563^\circ$ & 0.933 & $1.552^\circ$ &0.970 & $1.036^\circ$&0.971\\
B: MASK+RIV+SAD & $0.676^\circ$ & {\bf 0.974} & $0.952^\circ$ & 0.969 &$0.559^\circ$ & 0.973 & $\bm{0.706^\circ}$ &0.975 &$\bm{0.485^\circ}$&{\bf 0.978}\\
C: MASK+RIV+SAD with aug. &$\bm{0.528^\circ}$ & 0.973 & $\bm{0.417^\circ}$ & {\bf 0.976} & $\bm{0.470^\circ}$& {\bf 0.974} & $0.722^\circ$ &0.977&$0.503^\circ$&0.966 \\ \hline
\end{tabular}
\label{tb:result}
\end{table*}
\subsection{Loss Function}

For the loss function of DOA estimation, we used the MAE, and for that of SAD estimation, we used the binary cross entropy (BCE) for estimating a time series of the probability of the target sound activity $\bm{a} = (a_1, ..., a_T)^{\top}$. To train the DNNs, we used the sum of these loss functions and simultaneously trained all networks in an end-to-end manner. Since DOA is a phase variable, the difference in the estimate and label of DOAs must be less than $\pi$. To guarantee this, we define the DOA loss for elevation and azimuth as
\begin{equation}
    \begin{split}
        \Delta\theta_t &= |\theta_t - \hat{\theta_t}|,\\
        \Delta\phi_t &= {\rm min}\left(|\phi_t - \hat{\phi_t}|,|(\phi_t\pm{2\pi} - \hat{\phi_t}|\right), 
    \end{split}
\end{equation}
respectively. Since DOA loss cannot be defined at time frames, which do not have target sources, it is only calculated in which the activity ground truth $z_t =1$. Therefore, by defining $\mathcal{Z} = \sum_{t=1}^T z_t$, the loss function is expressed as
\begin{align}
    \mathcal{L} &=\mathcal{L}^{doa} + \mathcal{L}^{sad} \nonumber \\
    &=\frac{1}{\mathcal{Z}}\sum_{t=1}^{T} z_t \left(\Delta\theta_t + \Delta\phi_t\right) 
    + \frac{1}{T} \sum_{t=1}^{T} \mbox{BCE}\left(z_t, a_t \right),
\end{align}
where $\mbox{BCE} (a, b)$ is the BCE of $a$ and $b$.

\section{EXPERIMENTS}
\subsection{Experimental Setup}
We conducted experiments using 200 FOA recordings without overlap of sound sources in the data set of TAU Spatial Sound Events 2019~\cite{SELD-DOA2}. The 200 sound recordings consisted of 1 to 4 splits. Since the training data are limited, we did not set up validation splits, and all data were divided into 4-fold combinations of 150 training datasets and 50 test datasets.

The proposed method was compared with the conventional IV-based method described in Section \ref{sec:iv_based} and a DNN-based DOA estimation method (hearafter, called as DNN-conv). For fair comparison, the DNN architecture of DNN-conv was almost the same as the proposed method. The DNN of DNN-conv consisted of two CRNNs. The first CRNN was used for directly estimating the azimuth and elevation, and whose components are almost same as those of RIVnet besides the number of output units. The second CRNN was used for estimating the activation label $a_t$, and whose components are almost the same as those of the MASKnet and SADnet was used. Since DNN-conv does not use a T-F mask, the CRNNs do not have a fully connected layer for mask estimation. The total number of trainable parameters of DNN-conv was 2.72 M, almost the same as that of the proposed method (2.79 M).

In all experiments, the sampling frequency was 48 kHz. For the STFT, an 8192-point Hanning window with 20-ms shift was used. The number of Mel-filterbanks applied to the spectrogram and IVs was set to 96. We fixed the learning rate for the initial 50 epochs and reduced it linearly between 50--100 epochs down to a factor of 100 using the ADAM optimizer, where we started with a learning rate of 0.001. We always concluded training after 100 epochs.

We used hard thresholding for the final decision of SADs: when probability $a_t$ exceeded the threshold $\alpha$, we determined the time-frame $t$ including an active source. We used $\alpha=0.5$. Since test datasets are known to have sound sources in 10$^\circ$ steps, the obtained DOAs were discretized at 10$^\circ$ intervals. Furthermore, for smoothing, the median value for DOA in the event was taken as the DOA of that event:
\begin{equation}
    \begin{split}
        {\rm DOA_{dis}} &= {\rm round}({\rm DOA}/10^\circ)*10^\circ, \\
        {\rm DOA_{med}} &= {\rm median}({\rm DOA_{dis}}
        [\tau_1 : \tau_2]
        ),
    \end{split}
    \label{eq:pps}
\end{equation}
where $\tau_1$ and $\tau_2$ are the onset and offset time, which are the event intervals derived using the activity prediction $a_t$.

\subsection{Results}
We conducted these experiments using DOA Error (DE) and frame-recall (FR) as metrics \cite{SELD-DOA2}. DE represents the error of the estimated angle, and FR represents the accuracy rate of activity detection. 

To confirm the effectiveness of RIVnet and MASKnet, we tested three architectures. The first architecture (A) did not have RIVnet, the second (B) had both RIVnet and MASKnet, and the third (C) was trained using both RIVnet and MASKnet using the \textit{16-pattern method} of \textit{FOA-domain spatial augmentation} \cite{FOA_aug}. Figure \ref{fig:DOAresult} compares DOA estimation using architecture C without using post-processing (\ref{eq:pps})  with conventional methods. The DOA estimation result of each time-frames of the proposed method is clearly closer to the DOA labels than those of the conventional methods. These results indicate that the accuracy of time-series DOA estimation improved with the proposed method. Table \ref{tb:result} lists the experimental results. The results indicate that the DEs of architectures B and C were always lower than that of A and that using dereverberation is effective for IV-based DOA estimation. In addition, the average DE of architecture C was lower than that of architecture B, which indicates that using FOA-domain spatial augmentation is effective for DNN-based DOA estimation, as reported in a previous study \cite{FOA_aug}. Moreover, the average DE and FR of architectures A--C were higher than that of the conventional methods. Thus, we conclude that the accuracy of parametric-based DOA estimation methods improve by combining DNNs for refining physical parameters. 

\section{CONCLUSION}
\label{sec:cncl}
We proposed a method of improving IV-based DOA estimation by denoising and dereverberation using DNN models. Through objective experiments on a single-source DOA estimation task, we confirmed that the proposed method outperformed a conventional IV-based DOA estimation method, and the average DE of the proposed method was $0.528^\circ$. Therefore, we conclude that denoising and dereverberation using DNNs are effective in improving IV-based DOA estimation.

\newpage


\end{sloppy}
\end{document}